\documentclass[12pt]{article}
\usepackage{amsmath}
\usepackage{amssymb}
\usepackage{epsfig}
\usepackage{cite}
\usepackage{color,colordvi}

\newcommand{\eq}{\begin{equation}}
\newcommand{\eqx}{\end{equation}}
\newcommand{\eqn}{\begin{eqnarray}}
\newcommand{\bi}{\begin{itemize}}
\newcommand{\eqnx}{\end{eqnarray}}
\newcommand{\ei}{\end{itemize}}

\newcounter{hran}

\def\MSbar{\relax\ifmmode\overline{\rm MS}\else{$\overline{\rm MS}${ }}\fi}

\begin{document}

\begin{center}
\hfill CERN-PH-TH/2010-069\\[-2mm]
\hfill IFT-UAM/CSIC-10-25\\

{\Large\bf Species and Strings }

\vspace{7mm}

\end{center}

\begin{center}

{\bf Gia Dvali}$^{a,b,d,c}$ and {\bf Cesar Gomez}$^{e}$

\vspace{.6truecm}

\vspace{.2truecm}

{\em $^a$Arnold Sommerfeld Center for Theoretical Physics\\
Department f\"ur Physik, Ludwig-Maximilians-Universit\"at M\"unchen\\
Theresienstr.~37, 80333 M\"unchen, Germany}


{\em $^b$Max-Planck-Institut f\"ur Physik\\
F\"ohringer Ring 6, 80805 M\"unchen, Germany}


{\em $^c$CERN,
Theory Division\\
1211 Geneva 23, Switzerland}


{\em $^d$CCPP,
Department of Physics, New York University\\
4 Washington Place, New York, NY 10003, USA}

\vspace{.2truecm}

{\em $^e$
Instituto de F\'{\i}sica Te\'orica UAM-CSIC, C-XVI \\
Universidad Aut\'onoma de Madrid,
Cantoblanco, 28049 Madrid, Spain}\\

\end{center}

\vskip .35in

\centerline{\bf Abstract}
\vskip .1in
 Based on well-known properties of semi-classical black holes,  we  show that weakly-coupled string theory can be viewed as a theory of $N \, = \, 1/g_s^2$ particle species. 
  This statement is a string theoretic realization of the fact that the  fundamental scale in  any consistent $D$-dimensional theory of gravity is not the Planck length $l_D$, but rather the species scale  $L_N \, = \, N^{1 \over D-2} \, l_D$.   Using this fact,  we derive the bound on semi-classical  black hole entropy in any consistent theory of gravity as $S \, > \, N$, which when applied to string theory provides additional evidence for the 
former  relation.   This counting also shows that the Bekenstein-Hawking entropy can be viewed as the entanglement entropy, without encountering any puzzle of species.   We demonstrate that the counting of species extends to the $M$-theory limit.   The role of the  species scale is now played by the eleven-dimensional Planck length,  beyond which 
resolution of  distances is {\it gravitationally-impossible}. 
 The conclusion is, that string theory is a theory of species and gets replaced by a pure gravitational theory in the limit when species become strongly coupled and decouple.

\noindent

\vskip .4in
\noindent


\section{The Species Scale} 

  Consider a  consistent  quantum field theory in $D \, = \, 4\, +d\, $ space-time dimensions in  which gravity flows to the Einsteinian theory in deep infra-red (IR).  This condition implies 
that at large distances the gravitational force is mediated by a  $D$-dimensional 
massless particle of spin-2, the graviton $h_{\mu\nu}$.   The corresponding Newton's 
constant we shall denote by  $G_{D} \, \equiv \, M_{D}^{-(D-2)}$, where 
$M_{D}$ is a $D$-dimensional Planck mass. 
  Let  $L_*$ be a length scale defining the gravitational ultra-violet (UV) cutoff of the theory, beyond which 
the quantum gravitational effects can no longer be ignored.  In particular, $L_*$ marks the 
size (generalized  Schwarzschild radius) of a smallest possible semi-classical black hole (BH).
 We shall assume,  that around and below the scale $M_*$ there are $N$ different elementary particle species.  To be more precise,  $N$  counts all the  {\it weakly-coupled}  elementary particles
 with the decay width  much less than their mass.
 
   Previous studies \cite{bound, us, ramygabriele}  have shown, that in any consistent theory satisfying the above conditions 
there is an absolute lower bound on the scale $L_*$, given by, 
\begin{equation}
L_*  \, > \,  L_N \, \equiv \, N^{1 \over D-2} \, l_D \, ,
\label{bound}
\end{equation}
where, $l_D \equiv M_D^{-1}$ is the $D$-dimensional Planck length.  

Below we shall refer to $L_N$ as the {\it species scale}. The aim of the present paper is to 
show that $L_N$ plays a fundamental role in any consistent theory of gravity.  Namely, $L_N$ is an absolute shortest distance, beyond which resolution of species is impossible, {\it 
in principle}. 
  In particular, gravitational consistency of string theory demands that $L_N$ sets the larger of the 
two scales,  the string length, $l_s$, or $l_D$. As we shall see,   in a weakly coupled string theory we have
$l_s \, = \, L_N$, whereas in the strongly coupled $M$-theory limit,  in which  $l_s \, \ll \, l_{11}$, the role of   $L_N$ is assumed by the Planck length, 
$L_N \, = \, l_{11}$.   

   We shall reproduce two different derivations of the bound (\ref{bound}), and show, that when applied to
string theory,  they reveal that string theory is a theory of species 
with their effective number given by, 
 \begin{equation}
  N_{eff} \, =\, {1 \over g_s^2} \, .
\label{keypoint}
\end{equation}
This may come as a surprise, since it is known 
that the number of string Regge excitations diverges exponentially with the mass level. 
 So one expects that from the effective field theoretic point of view at high-energies the string theory for all the practical purposes should behave as a theory of 
 exponentially large number of species.   Contrary to this expectation,  the effective number of species resolved by gravity  appears to be much smaller and finite.  This point of view was already suggested by the findings of \cite{dvalilust1}, which indicated that  effective number of string  resonances participating in the evaporation of semi-classical BHs is not exponentially large, and is limited by $1/g_s^2$. 
  Our  present analysis reinforces this conclusion, but without any reference to the details of  the BH and string dynamics in UV.   We shall rather show that it follows solely from the 
extremely well-understood properties of semi-classical BHs and the known  relations among the  scales in string theory. 

  The conclusion that will emerge from our analysis is,  that string theory is theory of species coupled 
  to gravity and this statement can be extended beyond the weak coupling.  In the strongly coupled limit of $M$-theory in which string species decouple, we are left with pure supergravity 
  theory in $11$ dimensions with a single propagating graviton multiplet. 
  In full accordance with (\ref{bound}),  the role of $L_N$ is then taken by the $11$-dimensional 
  Planck length $l_{11}$, which sets the absolute bound on the resolution capacity of any detector.

  In addition,  we derive the bound on entropy  in any theory of the above sort as\cite{ramy},
  \begin{equation}
  S \, > \, N \, .
  \label{entropybound}
  \end{equation}
  When applied to string theory this bound again translates as the relation (\ref{keypoint}).
 
   One of the direct consequences of our species counting, is the full equivalence between Bekenstein-Hawking and entanglement entropies without encountering any puzzle of species.

 \section{Bound on Species Scale from Black Hole Evaporation}
      
 We shall define the effective number of species using the properties of semi-classical black holes.  Our focus will be on classically-static, neutral, non-rotating BHs on asymptotically 
 flat $D$-dimensional space.    We  shall assume, that the BH in question is {\it by design} neutral under all possible long-range gauge fields, and  the only characteristic quantum number is its mass $M$, which determines its Schwarzschild radius $R_g(M)$.  Notice, that as shown in \cite{ramygabriele},  in any ghost-free 
 $D$-dimensional theory, for given $M$, $R_g$ is minimal in a pure-Einsteinian gravity, 
 in which the only propagating $D$-dimensional degree of freedom is massless spin-2. The relation between $R_g$ and $M$ is then given  by (we shall ignore the factors of order one throughout the paper),
 \begin{equation}
 R_g^{D-3}\, = \, M \, G_D\, .
 \label{bhradius} 
 \end{equation}
 Any departure from $D$-dimensional Einstein at short distances can only increase $R_g$ per given $M$.  This is because 
\cite{stronggia} the 
 addition of any extra gravitational degrees of freedom
 can only strengthen 
  the gravitational attraction in the classical domain.   Thus,  any consistent deviation from  the 
 Einstein gravity  increases the value of $R_g$ relative to Einstein.   Other than this constraint,  we shall keep the relation between $R_g$ 
 and $M$ completely general.  The crucial property, however, is that the Hawking temperature of a semi-classical BH is always $T \, = \, R_g^{-1}$.  
  
     Under the above conditions, the evaporation of a semi-classical black hole obeys the following law of the temperature-change, 
 \begin{equation}
  {1 \over T}  {dT \over T} \, = \,  \Gamma_{total} \, , 
  \label{totalrate}
  \end{equation}
  where $\Gamma_{total}$ is the total rate of the emission of all the particle species. 
 The evaporation rate into heavy species of mass $m$,  is suppressed by the Boltzmann factor
 e$^{-{m\over T}}$.
  
 So,  for defining the effective number of species at the given temperature, it is most useful to normalize the emission rate to that of a massless graviton.  The latter rate by general covariance is given by, 
 \begin{equation}
 \Gamma_{gr} \,  = \,   T 
\left ( {T \over M_D} \right )^{D-2} \, .
\label{grate}
\end{equation}
We can thus introduce the dependence on the number of species in BH evaporation rate in the following way,  
 \begin{equation}
  {1 \over T}  {dT \over T} \, = \,  \Gamma_{gr} \, N \, .
  \label{totalrate}
 \end{equation}
 Thus, $N$ counts the BH evaporation rate in the units of the graviton emission rate $\Gamma_{gr}$. 
  Using the explicit form of (\ref{grate}) for $\Gamma_{gr}$ we can rewrite (\ref{totalrate}) as,  
 \begin{equation}
 {1 \over T^2}  {dT \over T} \,  = \, 
\left ( {T \over M_D} \right )^{D-2} \, N \, .
\label{rate}
\end{equation}
The usefulness of the latter equation is that the left hand side of it represents a semi-classicality parameter (we shall call it $\xi$),  
 \begin{equation}
\xi(T) \, \equiv  \,  {1 \over T^2}  {dT \over T} \, , 
\label{xi}
\end{equation}
 which defines the semi-classicality condition of the BH  through the inequality, 
 \begin{equation}
\xi(T) \,  \ll  \,  1 \, .
\label{semicond}
\end{equation}
Any classically-stable static non-rotating neutral BH must satisfy the above condition. 
Violation of it means, that the BH is out of the semi-classical regime.  The lowest temperature  $T_*$
for which the condition (\ref{semicond}) is violated, 
marks the scale of the gravitational cutoff for the classical theory.  That is $T_* \, = \, M_*$. 
The corresponding BH size,  $T_*^{-1}$,  defines the shortest length-scale of semi-classical gravity.   
  The value of the cutoff  $M_*$ is thus uniquely determined by the value of the temperature 
  $T_*$ for which  $\xi(T_*) \,  = \, 1$.  This leads to the bound (\ref{bound}) as well as to a maximum BH temperature $T_* \, = \, 1/L_N$. As we will discuss in the next section, this bound on temperature naturally fits with the Hagedorn bound in string theory.

 Notice,  that although for our derivation it is enough that the Schwarzschild radius is given by the usual $D$-dimensional Einstein relation (\ref{rpixel}), the bound on $L_N$ is 
in fact absolute.  It cannot be avoided even if,  let us say, the  laws of classical gravity start to depart from $D$-dimensional Einstein's general relativity at short distances.   
 As explained above,  any modification of Einsteinian gravity in the weakly-coupled classical domain can only strengthen 
the gravitational interaction relative to Einstein \cite{stronggia}.   From here it follows \cite{ramygabriele}, that Einsteinian gravity has the smallest  horizon per given mass of a BH.  Thus, any hypothetical deviation of classical gravity 
from $D$-dimensional Einstein's theory,  can only increase  the species scale relative to the value of $L_N$ given by (\ref{bound}), but can never make it shorter.

 
 \section{Species Scale as the Holographic Scale}
 
  An alternative derivation of the species scale \cite{us} follows from the impossibility of resolving species 
  identities down to arbitrarily short length scales.  The fundamental obstacle is created by gravity, as we shall now explain. 
     Physically,  having $N$ distinct species means that one can label them, and distinguish 
 these labels by physical measurements.   Let $\Phi_j$ be particle species and 
  $ j \, = 1,2, ...N$ be their labels.   One can talk about species as long as their identities can be resolved, at least {\it in principle}.  Gravity puts an insuperable lower bound on the resolution 
 capacity of any detector.   To see this, let us assume  that  $L$  is the size of an elementary pixel of the detector that is decoding the species identities.  Elementary act of recognizing the 
 species label $j$ is a scattering process in which an unknown particle scatters off an elementary
 pixel.   In order to read the species label, the pixel must contain  either a  sample of all the $N$ different species, or an equivalent information.   This fact automatically limits the size of the pixel from below. 
 Indeed, localization of each sample particle within the pixel of size $L_{pixel}$ costs energy $E\, = \, 1/L_{pixel}$, which implies that the total mass within the pixel is at least $M_{pixel} \, = \, N/L_{pixel}$. 
The corresponding  Schwarzschild radius is \,  
\begin{equation}
  R_{pixel}^{D-3} \, = \, {N \over L_{pixel}} \, G_{D} \, .
  \label{rpixel}
  \end{equation}
 The key point is, that such a pixel cannot be made arbitrarily small,  since eventually its Schwarzschild radius will 
 exceed the size of the pixel,  and this will happen while the size is still larger than the Planck length $l_D$!  This means that at the crossover size,  the pixel will collapse into a 
 {\it  classical}  BH,  and the information will take much longer time (if ever) to be retrieved.
 Notice that this lower bound is absolutely insensitive to the question of the information loss by the BH. Regardless, whether information is lost or not, once the pixel collapses into a semi-classical BH, one has to wait at least the evaporation time, which in any case exceeds $L_N$.  
 
 Thus, for resolving the species identities on the scale $L_{pixel}$ it is necessary that $R_{pixel} \,  < \, L_{pixel}$.  This condition after taking into the account (\ref{rpixel}), gives  the limiting size of the pixel as  $L_N$ given by  (\ref{bound}).  
 
 The holographic \cite{thooft, SW} meaning of $L_N$ is actually quite transparent. In fact $L_N$ is the minimal size where we can store $N$ bits of information. The reason this holographic scale coincides with the species scale is,  that the physical resolution of species requires, as discussed above, at least a number of information bits equal to the number of species. The only assumption in this holographic argument is to identify , as it is customary in our present interpretation of holography, the gravitational Planck length with the holographic scale defined as the minimum size where a bit of information can be stored. The holographic bound on information storage implies that whenever we try to prove sub-Planckian scales by localizing a particle in a region of size $l$ smaller than $l_D$, the corresponding dual description of that sub-Planckian  physics requires a number of holographic degrees of freedom equal to $(\frac{l_D}{l})^{\frac{D-2}{D-3}}$. Gravity as the way to set the holographic scale leads to an IR/UV correspondence between the energy scales $\frac{1}{l}$ we want to probe and the energy $(\frac{l}{l_D^{D-2}})^{\frac{1}{D-3}}$ ($\frac{l}{l_{pl}^2}$ in four dimensions ) of the corresponding dual holographic degrees of freedom.

\section{Species Scale in String Theory} 

 \subsection{Breakdown of Black Hole Semi-Classicality in Ten Dimensions} 
  
  We shall now show that the above facts lead us to the relation (\ref{keypoint}) in string theory. 
 The first immediate evidence for this relation comes from applying (\ref{rate}) to the evaporation of the semi-classical BHs in ten dimensions.  This gives,  
  \begin{equation}
 \xi(T) \,  = \, 
\left ( {T \over M_{10}} \right )^8 \, N \, .
\label{rate10}
\end{equation}
Assuming that the BH evaporation stays semi-classical (implying $\xi(T) \, \ll \, 1$)  at least till temperatures of order 
$T \lesssim \, M_s$, we get the bound 
\begin{equation}
  N  \, < \, (M_{10} /M_s)^8\, ,
  \end{equation}
  which after substituting the well known relation between the string and the Planck scales,
  \begin{equation}
  M_{10}^8 \, = \, {M_s^8 \over g_s^2} \,, 
  \label{stringscale}
  \end{equation}
  gives
  (\ref{keypoint}). 
  
    We see,  that the requirement that ten-dimensional BHs should ``afford" to stay in semi-classical regime till the temperature $l_s$,  immediately implies the bound (\ref{keypoint}) on the number of species  in string theory. 
  

  \subsection{Chan Paton  Factors: $D$-Branes as Species-Resolving Device} 
  
       The same conclusion  arises from considering  the Chan Paton (CP) factors.   These factors count the independent end-points of the open strings attached to $D$-branes. In this way   CP factors  label  different zero mode species in the world volume theory of the $D$-branes. 
  A stuck of $n$ $D$ branes  results in $n^2$ CP factors and thus in $N \, = \, n^2$ zero mode species (to be more precise in $n^2$ super-species, the degeneracy factor within the supermultiplet will be ignored).    
 So,  naively it may seem that we can arbitrarily increase the number of species by piling up an arbitrarily large number of $D_9$-branes on top of each other. 
  However, the number of species one may obtain in this way is limited by (\ref{keypoint}). 
  The physical reason for this limit can be understood from the fact,  that  the species label is the simplest bit in which one can encode information, and $D$-branes are devices for decoding this information. 
  In other words a stuck of $n$  $D$-branes represents a device that can resolve  
  $N \, = \, n^2$ species.  One should be able to decode this information at least down to 
  the distances of order the string length  $l_s$. 
    On the other hand, as we know, the minimum size of a pixel with the resolution capacity 
    of $N$ species in any consistent theory of gravity 
    is bounded by the scale  $L_N$.   An immediate connection between $N$ and $g_s$ then follows from the fact that  we should be able to read CP factors at least down to the string length, and thus  $L_N$ should not  exceed $l_s$, which implies (\ref{keypoint}).     
 
      The gravitational reason for the limited capacity of $D$-branes to support 
      large number of species  has a clear  physical meaning.    
 Indeed,  there is an obvious gravitational obstacle on how many CP factors we can generate by increasing the number of $D$-branes.  This obstacle arises in the following way. 
   Each $D_9$ brane carries a tension $T_9\, = \, {M_s^{10} \over g_s}$.  So, the collective 
gravitational source created by a stuck of $n$ branes (the source of the ten-dimensional zero mode graviton) is 
\begin{equation}
R^{-2} \, = \, nT_9 G_{10} \, = \, ng_s \, M_s^2 \, .
\label{radius}
\end{equation}
Requiring that this quantity cannot exceed the string tension $M_s^2$, leads to (\ref{keypoint}) as the bound on the number of CP species.  

Note, that if instead we consider a stuck of $n$ $D_p$-branes,  for the gravitational radius in terms of the ten dimensional species scale, $L_{N}$, we get
\eq
R(p,N) = \lambda^{\frac{p-3}{4(7-p)}} L_{N} = \lambda^{\frac{1}{7-p}} l_s \, ,
\eqx
where  $\lambda$ is the t'Hooft coupling. 
Only for the special case of $D_3$-branes this gravitational IR cutoff coincides with the species scale $L_{N}$. This agreement is another manifestation of the IR/UV correspondence for $D_3$ branes which is a crucial ingredient for AdS/CFT 
correspondence \cite{ADSCFT}.  

   An interesting question is what is the role (if any) of supersymmetry and BPS properties of the 
   $D$-branes in the species count.   The reason why we are bringing up this question is,  that for the  BPS $D$-branes the world-volume metric is formally flat despite the existence of  the 
   gravitational tension.  So naively, one could add arbitrary number of branes, without encountering  an immediate  problem with the horizon, and thus,  with retrieving the information.  Thus, naively one may think that BPS properties avoid the holographic bound.
   Of course, this is not the case, since the evaporation rate of a semi-classical and 
   neutral (and thus, non-BPS)  BHs  to the leading order is not sensitive to the 
   supersymmetry properties of the asymptotic space. 
  Obviously, as long as the background is asymptotically flat, 
 such a non-BPS BH of size $\gg l_s$ can always be formed regardless  the supersymmetry properties 
 of the asymptotic background.   Then, it must evaporate democratically into all the species, according to law given by (\ref{rate10}).   This evaporation tells us, that if the stuck of $D$-branes violates the species bound (\ref{keypoint}), the BH must become non-semi-classical at distances much larger than the string length, which would be a clear inconsistency.  So, the bound (\ref{keypoint}) must be respected.  However,  an interesting  
 question is whether the  system responds by some other dynamical effect preventing 
 growth of CP-species beyond the bound (\ref{keypoint}), independent of the BH-argument. 
    
  \subsection{Regge  Species} 

  We can distinguish the two notions of species in string theory. These are  Regge excitations and CP-species, 
  and they contribute differently into the species count.  Although, formally the number of Regge resonances is exponentially large, they contribute at best the factor $1/g_s^2$ into the 
  species number. The reason is,  that only a small fraction of the  Regge resonances contributes into the evaporation of a semi-classical BH \cite{dvalilust1}. 
  In fact,  what we have discovered for the Regge excitations is,  that when the "species scale" evaluated with respect to the number of Regge excitations becomes equal to the string length, the string state itself becomes a black hole and  the counting is self-truncated. 
  
   As for CP-species, they represent the zero mode excitations and must contribute the same 
   rate as the zero mode graviton  into the evaporation of any semi-classical BH. However, the number of CP-species cannot be increased arbitrarily for free, since they require existence of $n \, = \, \sqrt{N}$ background $D$-branes as their sources.  But, the latter number is automatically limited by the requirement, that the gravitational tension of the  system must not exceed the string scale.  In other words, there is an inevitable conflict between increase of the CP-species and limiting their gravitational backreaction, which results in the bound on  the number of species given by (\ref{keypoint}).

   \section{Entropy Bound and the String-Black Hole Correspondence} 
  
   We shall now show that the scale $L_N$ implies the lower bound on the BH entropy,
   \begin{equation}
     S \, > \, N \, , 
     \label{etropy}
     \end{equation}
     in any consistent  $D$-dimensional theory of gravity.  The bound follows from the fact that 
     the scale $L_N$ on one hand is the size of a smallest possible semi-classical BH and on the other hand is the size of a smallest possible pixel that carries the information about the 
 $N$ particle species labels.  Notice, that the species label is a simplest form of  the information storage.   Assuming that the occupation number of species corresponding to each label is 
 $0$ or $1$, the number of possible states of such a pixel scales as $2^N$, which gives the minimal  entropy of the BH to which such a pixel can collapse being $S > N$. 
 
  This bound gives yet another evidence for (\ref{keypoint}).  Indeed,  when the black hole reaches the size of the species scale $L_N$ their entropy saturates the bound. On the other hand a string state becomes a black hole,  i.e., a state with gravitational radius equal to the string length and with the string entropy equal to the Bekenstein-Hawking entropy, for $S=\frac{1}{g_s^2}$ \cite{HP, LS}. Actually if we assume string theory as the UV completion,  we find $\frac{1}{g_s^2}$ as an absolute bound on the BH entropy reached when the BH becomes of string size. This bound on the BH entropy sets the bound on the BH size as the string length and the Hagedorn temperature as the corresponding maximum temperature. We can obviously map this string physics with the physics we get in any consistent theory of species and gravity. Namely the entropy bound $S>N$ maps into the stringy bound $S>\frac{1}{g_s^2}$ using (\ref{keypoint}). 
 In the other words, what happens to a black hole when it reaches the species scale is a Horowitz-Polchinski transition into a string state of a string theory with $l_s =L_N$ and $N=\frac{1}{g_s^2}$. 

 
  
    We  wish to briefly discuss now how our bound relates to a naive counting 
   of string Regge excitations.  In string theory we can divide species into two categories. These are, the zero modes, in particular given by the CP factors, and the string Regge resonances.   The  
   semi-classical BH evaporation argument in ten-dimensions, presented above,  puts the 
   absolute bound on the number of zero modes, since  all of these modes have to participate in the evaporation of semi-classical BH. 
    Once the BH reaches the string size, the Regge excitations set in.  The naive counting gives, that the number of  these excitations is exponentially large.  However, the BH evaporation and holographic 
    arguments indicate that the effective number of these species must be bounded 
 by (\ref{keypoint}).    In the other words, most of the string-theoretic  ``species" are not species 
 at all!  
     The reason for this is as follows. 
     
     First, the Regge excitations above the certain oscillator number are  simply equivalent to BHs, as it is indicated by string-BH correspondence.   
  At high energies the number of states in string theory grows with energy as $e^{l_{s}E}=e^{\sqrt{N_{osc}}}$, with $N_{osc}$ being the oscillator level. 
    This  leads to a stringy entropy $S\, \sim \, El_{s}  = \sqrt{N_{osc}}$ and therefore to a temperature $T\sim \frac{1}{l_{s}}$. Moreover, this temperature, the Hagedorn temperature, is an absolute bound. Now, for Einsteinian black holes the entropy scales with energy as $S\sim (E l_D)^{{D-2 \over D-3}}$ for $D$ the number of space-time dimensions. 
 The black hole becomes a string state when both entropies scale with the energy in the same way, i.e.  in ten-dimensions this happens when $El_{10}^8 = l_{s}^7$,  which after using the relation between the string and Planck lengths implies $\sqrt{N_{osc}}=\frac{1}{g_{s}^2}$. 
 Thus, any string state corresponding to a higher oscillator number is no longer a 
 particle, but rather a BH. 
 In other words, starting from the opposite end,   when the black hole reaches the species scale its energy is related to the species scale as the string energy is related to the string length,
 \begin{equation}
 M= N L_N^{-1} \, ,
 \end{equation}
 with the number of species playing the role of $\sqrt{N_{osc}}$.
That is, the BH in question  becomes a string state with string length being  the species scale and the number of species related to the string coupling by (\ref{keypoint}). Thus, in a theory of gravity the species scale fixes a maximal "Hagedorn" temperature equal to $\frac{1}{L_{N}}$ as well as the analogous "Hagedorn" transition.
 
  Now as for the   Regge excitations with the lower oscillator number, although their formal number is exponentially large, majority of these states cannot contribute into the evaporation of any BH that can be treated semi-classically. This was illustrated by the analysis of \cite{dvalilust1}. The reason is the suppression of the emission due to the size of a BH, 
which can also be connected  to Veneziano-type softening of the string interactions at short distances. 
  
   To summarize, the  reason why number of Regge resonances contributes at best 
   $\sim 1/g_s^2$ to the species count is,  that the majority of these resonances are either BHs 
   or are not produced in the evaporation of the semi-classical BHs.

  \section{Entanglement Entropy} 
  
    We now wish to show that the relation (\ref{keypoint}) independently arises from the interpretation of the Bekenstein-Hawking entropy ($S_{BH}$) as the entanglement entropy $S_{ent}$.  Let us  compute an entanglement entropy of a $D$-dimensional Einsteinian BH of area  $A$. As it is well known \cite{ent},  $S_{ent}$, just as $S_{BH}$, scales as area. But, unlike the latter entropy, $S_{ent}$ depends both on the cutoff as well as on the number of species, 
   \begin{equation}
     S_{ent} \, = \, A \, L_*^{2-D} \, N \, . 
   \label{Se}
   \end{equation}
  This dependence for some-time was a source of a seeming discrepancy that goes under the name of the {\it species puzzle}
(see \cite{puzzle} for a review).   In reality, however, this puzzle does nor exist \cite{solukhin}
 in the view of the bound (\ref{bound}). 
 
  The cutoff $L_*$ that must enter in the computation of $S_{ent}$ is exactly the species scale $L_N$, since beyond this
  scale species cannot be resolved {\it in principle}. 
   Once this is taken into the account the discrepancy between the scalings of the two entropies disappears. 
   Let us now apply this reasoning to the ten dimensional BHs.   Eq (\ref{Se}) can then be rewritten in the following way, 
     \begin{equation}
     S_{ent} \, = \, {A \over L_N^8} \,  \, N \, .
   \label{Se1}
   \end{equation}
   Demanding that $L_N\, = \, l_s$ and equating the resulting $S_{ent}$ to the $S_{BH}$, we get 
  (\ref{keypoint}). 
   
     In fact,  we can turn the latter result around and use (\ref{keypoint}), which is evident from the independent arguments presented earlier in this paper, to show that the entanglement entropy computed with the obvious  UV cutoff,  $l_s$, exactly reproduces the Bekenstein-Hawking entropy in ten-dimensions.
       
Moreover, for any local quantum field theory with $N$ elementary species the UV cutoff dependent part of the entanglement entropy can be always expressed as, 
\eq
S_{ent} = S_{BH} \left (\frac{L_N}{L_{UV}} \right )^2 \, , 
\eqx
with $S_{BH}$ the entropy of a black hole with the horizon area equal to the boundary of the region we are tracing, and $L_{UV}$ the UV cutoff length of the theory without gravity.

If we now assume,  in accordance with  the holographic principle,  that the BH entropy is the maximum entropy we can associate with a given region, we could conclude that  $L_{UV}$ should be necessarily bounded by the species scale.
In other words,
we could conclude that if a local quantum field theory with $N$ elementary species can be consistently coupled to gravity,  then necessarily this local quantum field theory must posses an UV completion that can be identified , in the large $N$ limit, with a string theory.  Moreover, since we can always define a gravity decoupling limit,  
\begin{equation}
N\, \rightarrow \, \infty, , \, \, \, \, \,  M_D \, \rightarrow \, \infty  \, 
\label{decoupling}
\end{equation}
 keeping the UV scale $L_N$ finite, we can tentatively interpret this argument as an indirect way to point out to the dynamical generation of a (string)  scale in the limit of infinite number of species.

   
   \section{Species Scale in $M$-Theory}

 We shall now ask, how does the species-count extend beyond the weakly-coupled string theory?
 Of course,  whenever a dual weakly-coupled description is known, the counting 
 translates trivially in terms of the bound on the  number of (weakly-coupled) species in that new  description. 
 We shall focus instead on a well known example in which no dual weakly coupled description is 
 known, and instead one has to go to $M$-theory. 
  
   As it is well known  \cite{Mtheory},  the strong coupling limit of type $II A$ string theory  takes us to $M$ theory, with the low energy description being $11$-dimensional gravity theory. The $11$-th dimension opens up in the limit $g_s \, \rightarrow \, \infty$ in which $D_0$-branes become massless and 
   give rise  to the perturbative states, which from the point of view of the low energy theory are Kaluza-Klein (KK) excitations of $11$-dimensional gravity compactified to $10$-dimensions.  
 We shall now show, that the species-counting given by (\ref{bound}) again works, but the difference is, that the scale $L_N$ is now determined by the $11$-dimensional Planck length. It is now  $L_{11}$ and not $l_s$ that sets the absolute 
 limit  on the resolution capacity of any detector.  
 
With $R$ denoting the radius of $11$-th dimension, the basic defining relations of M-theory are:
\eq\label{one}
R\, = \, g_{s}l_{s}
\eqx
and
\eq
M_{10}^8 = R M_{11}^9 \, ,
\label{massscales}
\eqx
where the first relation follows from identifying the $D_0$-branes (with mass $g_sl_s$) with the 
first KK excitations of mass $1/R$.   The equation  (\ref{massscales}) is the standard geometric relation between the $10$ and $11$-dimensional Planck scales.  
Equations (\ref{one}) and (\ref{massscales}),  together with the basic relation  $l_{10} =g_{s}^{\frac{1}{4}} l_{s}$, lead to the following  relation between the string scale and the eleven dimensional Planck length, 
\eq\label{three}
l_{11} =g_{s}^{\frac{1}{3}} l_{s} \, .
\eqx
The above equations are already enough to understand how the bound (\ref{bound}) is 
saturated in the $M$-theory limit.  Indeed,  for $g_s \, \rightarrow \, \infty$, string excitations 
become strongly coupled and no longer fall under our definition of perturbative species. 
Instead, the role of the perturbative states is fulfilled by the KK excitations that originate 
from $D_0$-branes.  Their number is given by $N \, =\, M_{11}R$.  Taking this into the account, we realize that the equation (\ref{massscales}) is nothing else but the saturation of the bound
(\ref{bound}), where the role of $L_N$ is played by $L_{11}$.  The latter fact has a clear physical meaning, since from equation (\ref{three}) it follows that $l_s$ is much shorter than $l_{11}$. 
Thus, one hits the strong gravity scale way before there is any chance to probe the string length physics.  As a result it is $L_{11}$ and not $l_s$ that plays the role of the species scale. 
 In fact, physics at distances shorter than $L_{11}$ cannot be resolved, in principle.

  In order to follow how the $L_N$ interpolates between $l_s$ and $L_{11}$ in two different descriptions, let us evaluate some useful relations for the two regimes.  
 Let us start with a weakly-coupled description and consider $n$ $D_0$ branes in $D=10$. The mass of this stuck is
\eq
M \, = \, \frac{n}{g_s l_s}
\eqx
The corresponding gravitational radius in $10 D$ is
\eq\label{two}
R_{g}^{7} = (g_s n)l_s^7
\eqx
that leads to the standard bound,
\eq
g_{s} <\frac{1}{n} \, ,
\label{gsandn} 
\eqx
as long as we require $R_{g}$ to be smaller than the string length.
Noticing that $D_0$ brane CP factors in $10 D$ create $N=n^2$ different types of stringy species in the string weak coupling regime,  the latter equation  reproduces the 
bound (\ref{keypoint}). 
 
 For strong string coupling the gravitational radius of $n$ $D_0$ branes will be bigger than the string length, meaning that the CP factors can no longer be resolved at the string length scale.  
 This is the gravitational reason why string excitations can no longer be considered as peturbative species.   

Let us consider now a graviton in eleven dimensions with the $11$th component of the momentum being $P_{11}=\frac{N}{R}$. 
The corresponding gravitational radius in eleven dimensions is
\eq
R_{g}^8 = l_{11}^9 \frac{N}{R} \, .
\eqx
Notice,  that for any $N \, < \, (RM_{11})$, this gravitational radius is $R_g \, < \, l_{11}$. 
The value of $N$ that saturates this inequality has a clear physical meaning.  From the point of view of an $11$-dimensional observer this is a maximal value of the graviton momentum 
before the latter collapses into a BH.  From the point of view of a $10$-dimensional observer 
this is a maximal number of species compatible with the bound (\ref{bound}).  The reason why the role of $L_N$ is played by $l_{11}$ is also clear, since this is the energy scale above which the Compton wave-length of the 
$11$-dimensional graviton crosses over below  its Schwarzschild horizon. 

The counting of species matches from both points of view. This agrees with the idea that  the eleven dimensional graviton is holographicaly described by $N$ $D_0$ branes. This was the M(atrix) conjecture \cite{Banks}).

What happens when we push $g_{s}$ into the strong coupling regime ?
Using the standard KK definition of the number of species
$n$ as 
\eq
n=\frac{R}{l_{11}}
\eqx
we get using (\ref{one}) and (\ref{three})
\eq\label{five}
n=g_s^{\frac{2}{3}} \, .
\eqx
Moreover from (\ref{three}) we get

\eq
l_{11} =g_{s}^\frac{1}{12} l_{10} \, ,
\label{l11l10}
\eqx
for $l_{10}$,  the ten dimensional Planck length.
This together with (\ref{five}) leads to
\eq
l_{11}=n^{\frac{1}{8}}l_{10}
\label{nspecies}
\eqx
that is the standard definition of the species scale in 10 D.

 To summarize, the following picture emerges.  In the weak string-coupling limit, 
 ten-dimensional species correspond to the  string vibration modes, and the role of 
 $L_N$ is played by the string scale $l_s$.  The latter scale is the first barrier 
 that one encounters in trying to resolve the short distance scales, way before reaching the Planck length.  Saturation of the bound (\ref{bound}) then implies (\ref{keypoint}). 
  
  In the strongly coupled limit,  string excitations can no longer be regarded as perturbative states.  Instead,  the role of the latter states is played by KK species of the compact $11$-th dimension.  The role of the species scale now is played by $l_{11}$, since this is the largest length scale beyond  which  resolution of states is {\it gravitationally impossible}. 
 The bound (\ref{bound}) then  emerges as the well-known geometric relation (\ref{massscales}).

\subsection{Formulation in Terms of the Dual Coupling}

We can present the bound (\ref{keypoint})
in completely general terms, both in the weak coupling and in the strong coupling regimes, provided in the strong coupling regime we replace $g_s$ by its corresponding dual,
 \begin{equation}
 g_s \, \rightarrow \, g_d \, .
\label{dualcoupling}
\end{equation}  
 We wish to illustrate this on the example of $M$-theory. 
As discussed above, in the weak coupling regime of ten-dimensional string theory 
the number of species is defined by the CP factors attached to $n$ $D_9$-branes,  i.e. $N=n^2$, which according to (\ref{gsandn}) is bounded by $1/g_s^2$. When we move to the strong coupling limit,  the dual light degrees of freedom become the $n$ $D_0$-branes and  we should expect that the number of species is now $N=n$, as it indeed is the case ( see equation (\ref{nspecies})). If we think in terms of a dual coupling ($g_{d}$) we should generalize the bound (\ref{keypoint})  in the strong coupling regime as 
\eq
N_{eff} \,  = \, \frac{1}{g_{d}^2} \, .
\label{dualbound}
\eqx
But then,  according to (\ref{five}) we must conclude that the dual coupling is
\eq\label{seven}
g_{d}= g_{s}^{-\frac{1}{3}} \, .
\eqx
The latter expression for the dual coupling is easy to understand using the dynamics of $D_0$-branes. In fact, the dynamics of $D_0$ branes is the $0+1$ - dimensional 
quantum mechanics obtained by the dimensional reduction of ten-dimensional Yang Mills, i.e. ( for the bosonic components ),
\eq
\int dt \frac{1}{g_{s}} Tr F_{\mu,\nu} F^{\mu,\nu} \, . 
\eqx
This gives the following interaction term for the scalar components $\phi_{i}$ $i= 1,,,9$,
\eq\label{eight}
\frac{1}{g_{s}} |\phi_{i} \times \phi_{j}|^2 \, .
\eqx
Now, since $l_{11}= g_{s}^{\frac{1}{3}} l_{s}$, we can move from string units into M-theory units by  
replacing,
\eq
\phi_{i} \rightarrow g_{s}^{\frac{1}{3}} \phi_{i} \, ,
\eqx
and transforming (\ref{eight}) into
\eq
g_{s}^{\frac{1}{3}} |\phi_{i} \times \phi_{j}|^2 \, .
\eqx
Reading the above expression as the definition of  the dual coupling,
\eq
\frac{1}{g_{d}} |\phi_{i} \times \phi_{j}|^2 \, ,
\eqx
leads to the desired relation (\ref{seven}) and to the species bound in strong coupling regime defined by 
(\ref{dualbound}) 
for $N=n$, the number of $D_0$-branes. 


\section{Final Comments}
In this note we have collected some evidence in the direction of mapping any consistent theory of gravity and $N$ weakly coupled species into a string theory. A theory of species is characterized by two fundamental lengths scales; the Planck length setting the intensity of gravity and the species scale given by  (\ref{bound})  that sets the minimal length where resolution of species is physically possible. The ultimate reason for the existence of  these two scales is the holographic meaning of gravity and the information content of species. Non BPS black holes with negative specific heat are the natural candidates to probe physics near the species scale. We have observed that at this scale black holes behave as string states for a weakly coupled string theory under the correspondence:
$l_s \rightarrow L_N$, $N \rightarrow \frac{1}{g_s^2}$. In this setup the weakly coupled string theory completes in the UV the theory of species, filling the gap between $L_N$ and the Planck scale. However, consistency of this interpretation of string theory at strong string coupling forces the existence of a weakly coupled dual description with the number of effective dual species again related to the dual string coupling by the same fundamental relation. We have presented some evidence that this is in fact the case for the M-theory dual description of type IIA strings. In summary, we have attempted to map string theory as a theory of two length scales, the string scale and the Planck scale, into its most natural cousin a theory with species coupled to gravity. Hopefully this correspondence could shed some novel light in order to foresee what is string theory.

  \vspace{5mm}
\centerline{\bf Acknowledgments}

This work is supported in part by  European Commission  under ERC advanced grant 226371 via CERN. 
The work of G.D is supported in part by Humboldt Foundation under Alexander von Humboldt Professorship,  by European Commission  under 
the ERC advanced grant 226371,  by  David and Lucile  Packard Foundation Fellowship for  Science and Engineering and  by the NSF grant PHY-0758032. 
The work of C.G. has been partially supported by the Spanish
DGI contract FPA2003-02877 and the C.M grant HEPHACOS
S2009ESP-1473.


\begin{thebibliography}{99}

\bibitem{thooft} 

G.~'t Hooft,
"Dimensional reduction in quantum gravity", gr-qc/9310026;
L.~Susskind,
"The World As A Hologram", J.\ Math.\ Phys.\  {\bf 36}, 6377 (1995),  hep-th/9409089.



\bibitem{bound}

G.~Dvali, ``Black Holes and Large N Species Solution to the
Hierarchy Problem,'' arXiv:0706.2050 [hep-th];

G.~Dvali and M.~Redi, ``Black Hole Bound on the Number of Species and Quantum Gravity at LHC,''
Phys. Rev.  {\bf D77} ( 2008) 045027,  arXiv:0710.4344 [hep-th];


G.~Dvali, ``Nature of Microscopic Black Holes and Gravity in Theories with Particle Species'',  arXiv:0806.3801 [hep-th];`


\bibitem{us} G.~Dvali and C~Gomez, ``Quantum Information and Gravity Cutoff in Theories with Species'',
Phys. Lett. {\bf B674}  (2009) 303,  arXiv:0812.1940 [hep-th] .

G.~Dvali and C.~ Gomez, 
 ``Strong Coupling Holography",  arXiv:0907.3237 [hep-th]

\bibitem{ramygabriele} R.~Brustein, G.~Dvali and G.~Veneziano, `` A Bound on Effective 
Gravitational Coupling from Semiclassical Black Holes",  
 JHEP 0910:085,2009,  arXiv:0907.5516 [hep-th]




\bibitem{ADSCFT}

 
J. M.~ Maldacena, 
 `` The Large N limit of superconformal field theories and supergravity", 
 Adv.Theor.Math.Phys.  {\bf 2 } (1998) 231, Int.J.Theor.Phys. {\bf 38}  (1999) 1113,  hep-th/9711200; 
 
  
S.S.~Gubser, I.R.~ Klebanov, A. M.~Polyakov,    
``Gauge theory correlators from noncritical string theory",  Phys. Lett. {\bf B428} (1998) 105, 
hep-th/9802109;
E.~ Witten,     ``Anti-de Sitter space and holography", Adv.Theor.Math.Phys. {\bf 2} (1998) 253,
hep-th/9802150





\bibitem{SW}

L.~ Susskind and  E.~ Witten, ``The Holographic bound in anti-de Sitter space", 
 arXiv:9805114 [hep-th]  




 \bibitem{dvalilust1}
 
 G.~Dvali and D.~Lust, 
``Evaporation of Microscopic Black Holes in String Theory and the Bound on Species",
arXiv:0912.3167 [hep-th] 

 \bibitem{ent}

W.~Israel, Phys. Lett. A 57 (1976) 107;

L.Bombelli, R.K. Koul, J.H. Lee and  R.D.~Sorkin,  Phys. Rev. D34, 373 (1986);

M. Srednicki, Phys. Rev. Lett. 71, 666 (1993); 

V.P. Frolov and I. Novikov, Phys. Rev. D48, 4545  (1993). 


\bibitem{puzzle}

J.D. Bekenstein,  ``Do we understand black hole entropy?", arXiv:gr-qc/9409015. 



  
  \bibitem{solukhin}
  G. Dvali and S. N. Solodukhin,  `` Black Hole Entropy and Gravity Cutoff", 
hep-th/08063976 


\bibitem{stronggia} 
 G.~ Dvali,  
`` Predictive Power of Strong Coupling in Theories with Large Distance Modified Gravity",
 New J.Phys.  {\bf 8}  (2006) 326,  hep-th/0610013;  

G.~Dvali,  O.~ Pujolas and  M.~ Redi
``Non Pauli-Fierz Massive Gravitons",   Phys. Rev. Lett. 101:171303, 2008.  arXiv:0806.3762 [hep-th]

\bibitem{ramy} 
 While this paper was in preparation, we became aware of  the following work that discusses the same bound from a different perspective. We thank the authors for sharing their preliminary results with us.

R.~Brustein and A.J.M. Medved,
``Bounds on Black Hole Entropy in Unitary Theories of Gravity", 
arXiv:1003.2850 [hep-th]
 


\bibitem{HP}

G. T. Horowitz and J. Polchinski, A correspondence principle for black holes and 
strings, Phys. Rev. D 55, 6189 (1997), hep-th/9612146 

G.T. Horowitz and J. Polchinski, Phys. Rev. D 57, 2557 (1998) [hep-th/9707170]; 

T. Damour and G. Veneziano, Nucl. Phys. B 568, 93 (2000) [hep-th/9907030].
\bibitem{LS}

L. Susskind, hep-th/9309145, 

E. Halyo, A. Ra jaraman, and L. Susskind, hep-th/9605112,

 E. Halyo, B. Kol, A. Ra jaraman, and L. Susskind, hep-th/9609075.


 
 \bibitem{Mtheory}
 
 M.J. Duff ,, P. Howe, T. Inami, K.S. Stelle, �Superstrings in D=10 from supermem- 
branes in D=11�, Phys. Lett. B191 (1987) 70; 

M.J. Duff , J. X. Lu, �Duality rotations in membrane theory�, Nucl. Phys. B347 (1990) 
394 ; 

M.J. Duff,  R. Minasian, James T. Liu, �Eleven-dimensional origin of string/string 
duality: a one-loop test�, Nucl. Phys. B452 (1995) 261 ; 

C. Hull and P. K. Townsend, �Unity of superstring dualities�, Nucl. Phys. B438 
(1995) 109, hep-th/9410167 ; 

P. K. Townsend, �The Eleven-Dimensional Supermembrane Revisited� Phys. Lett. 
B350 (1995) 184 hep-th/9501068 ; 

�String-membrane duality in seven dimensions�, Phys. Lett. 354B (1995) 247, hep- 
th/9504095 ; 

C. Hull and P. K. Townsend, �Enhanced gauge symmetries in superstring theories�, 
Nucl. Phys. B451 (1995) 525, hep-th/9505073; 

E. Witten �String Theory Dynamics in Various Dimensions�, Nucl.Phys. B443 (1995) 
85 hep-th/9503124. 

\bibitem{Banks}
T.~Banks , W. ~Fischler , S.H.~ Shenker, L.~ Susskind, `` M theory as a matrix model: A Conjecture", 
Phys.Rev.D55:5112-5128,1997. hep-th/9610043




 \end{thebibliography}
\end{document}